\documentclass[conference]{IEEEtran}

\renewcommand\IEEEkeywordsname{Index Terms}
\usepackage{graphicx}
\usepackage{subcaption}
\usepackage{tikz,xparse}
\usetikzlibrary{dsp,chains}
\usetikzlibrary{matrix}
\usetikzlibrary{spy}
\usepackage{mathptmx}
\usepackage{verbatim}
\usepackage{calc}
\usepackage{ifthen}
\usepackage{xifthen}
\usepackage{cancel}
\usepackage{bm}
\usepackage{verbatim}
\usepackage{multirow}
\usepackage{cite}

\usepackage[nolist]{acronym} 
\usepackage{pgfplots}
\usetikzlibrary{arrows,shapes,graphs,graphs.standard,quotes,decorations.markings}
\usetikzlibrary{matrix}
\pgfplotsset{compat=newest}

\usepackage[hyphens]{url}

\usepackage[bookmarks=false]{hyperref}
\usepackage{units}
\usepackage{amsmath, amsbsy, amssymb, latexsym }
\hypersetup{bookmarksdepth=-2}
\usepackage{comment}
\usepackage[utf8]{inputenc}
\usepackage{xcolor}
\usepackage{enumitem}
\usepackage{algorithm}
\usepackage{algpseudocode}

\usepackage{etoolbox}

\usepgfplotslibrary{external} 
\tikzset{>=latex}

\DeclareMathOperator*{\maxstar}{max^\ast}

\algnewcommand\algorithmicforeach{\textbf{for each}}
\algdef{S}[FOR]{ForEach}[1]{\algorithmicforeach\ #1\ \algorithmicdo}

\algnewcommand\algorithmicswitch{\textbf{switch}}
\algnewcommand\algorithmiccase{\textbf{case}}
\algnewcommand\algorithmicassert{\texttt{assert}}
\algnewcommand\Assert[1]{\State \algorithmicassert(#1)}%
\algdef{SE}[SWITCH]{Switch}{EndSwitch}[1]{\algorithmicswitch\ #1\ \algorithmicdo}{\algorithmicend\ \algorithmicswitch}%
\algdef{SE}[CASE]{Case}{EndCase}[1]{\algorithmiccase\ #1}{\algorithmicend\ \algorithmiccase}%
\algtext*{EndCase}%

\definecolor{mittelblau}{RGB}{0, 126, 198}
\definecolor{violettblau}{cmyk}{0.9, 0.6, 0, 0}
\definecolor{rot}{RGB}{238, 28 35}
\definecolor{apfelgruen}{RGB}{140, 198, 62}
\definecolor{gelb}{RGB}{255, 229, 0}
\definecolor{orange}{RGB}{244, 111, 33}
\definecolor{pink}{RGB}{237, 0, 140}
\definecolor{lila}{RGB}{128, 10, 145}
\definecolor{hellgrau}{RGB}{224, 224, 224}
\definecolor{mittelgrau}{RGB}{128, 128, 128}
\definecolor{dunkelgrau}{RGB}{80,80,80}
\definecolor{anthrazit}{RGB}{19, 31, 31}
\definecolor{darkgreen}{RGB}{34,139,34}
\colorlet{Mycolor1}{green!10!orange!90!}
\tikzset{
       vnd/.style={
        shape=circle,
        fill=black,
        draw,
        inner sep=0pt,
        minimum size=0.2cm},
        cnd/.style={
        shape=rectangle,
        fill=white,
        draw,
        minimum width=0.05mm,
        minimum height = 0.05mm}, 
         vndR/.style={
        shape=circle,
        fill=red,
        draw,
        inner sep=0pt,
        minimum size=0.2cm},
        cndR/.style={
        shape=rectangle,
        fill=white,
        draw=red,
        minimum width=0.05mm,
        minimum height = 0.05mm}
}

\IEEEoverridecommandlockouts

\renewcommand{\vec}[1]{\mathbf{#1}}

\newcommand{\yv}{\vec{y}}

\newcommand{\Gm}{\vec{G}}
\newcommand{\Hm}{\vec{H}}

\newcommand{\Lm}{\vec{L}}

\newcommand{\Rm}{\vec{R}}

\newcommand{\aposteriori}{{\em a posteriori} }

\begin{document}
	
\setlength{\abovedisplayskip}{-8pt}
\setlength{\belowdisplayskip}{2pt}

\begin{NoHyper}
\title{CRC-Aided Belief Propagation List Decoding of Polar Codes}

\author{\IEEEauthorblockN{Marvin Geiselhart, Ahmed Elkelesh, Moustafa Ebada, Sebastian Cammerer and Stephan ten Brink} \thanks{This work has been supported by DFG, Germany, under grant BR 3205/5-1.}
	\IEEEauthorblockA{
		Institute of Telecommunications, Pfaffenwaldring 47, University of  Stuttgart, 70569 Stuttgart, Germany 
		\\\{geiselhart,elkelesh,ebada,cammerer,tenbrink\}@inue.uni-stuttgart.de
	}
}

\makeatletter
\patchcmd{\@maketitle}
{\addvspace{0.5\baselineskip}\egroup}
{\addvspace{-0.6\baselineskip}\egroup}
{}
{}
\makeatother

\maketitle
\begin{acronym}
 \acro{ECC}{error-correcting code}
 \acro{HDD}{hard decision decoding}
 \acro{SDD}{soft decision decoding}
 \acro{ML}{maximum likelihood}
 \acro{GPU}{graphical processing unit}
 \acro{BP}{belief propagation}
 \acro{BPL}{belief propagation list}
 \acro{CA-BPL}{CRC-aided belief propagation list}
 \acro{LDPC}{low-density parity-check}
 \acro{HDPC}{high density parity-check}
 \acro{BER}{bit error rate}
 \acro{SNR}{signal-to-noise-ratio}
 \acro{BPSK}{binary phase shift keying}
 \acro{BCJR}{Bahl-Cocke-Jelinek-Raviv}
 \acro{AWGN}{additive white Gaussian noise}
 \acro{MSE}{mean squared error}
 \acro{LLR}{log-likelihood ratio}
 \acro{MAP}{maximum a posteriori}
 \acro{NE}{normalized error}
 \acro{BLER}{block error rate}
 \acro{PE}{processing element}
 \acro{SCL}{successive cancellation list}
 \acro{SC}{successive cancellation}
 \acro{BI-DMC}{Binary Input Discrete Memoryless Channel}
 \acro{CRC}{cyclic redundancy check}
 \acro{CA-SCL}{CRC-aided successive cancellation list}
 \acro{BEC}{Binary Erasure Channel}
 \acro{BSC}{Binary Symmetric Channel}
 \acro{BCH}{Bose-Chaudhuri-Hocquenghem}
 \acro{RM}{Reed--Muller}
 \acro{RS}{Reed-Solomon}
 \acro{SISO}{soft-in/soft-out}
 \acro{PSCL}{partitioned successive cancellation list}
 \acro{SPA}{sum product algorithm}
 \acro{LFSR}{linear feedback shift register}
 \acro{3GPP}{3rd Generation Partnership Project }
 \acro{eMBB}{enhanced Mobile Broadband}
 \acro{CN}{check node}
 \acro{VN}{variable node}
 \acro{PC}{parity-check}
 \acro{GenAlg}{Genetic Algorithm}
 \acro{AI}{Artificial Intelligence}
 \acro{MC}{Monte Carlo}
 \acro{CSI}{Channel State Information}
 \acro{FG}{factor graph}
 \acro{URLLC}{ultra-reliable low-latency communications}
 \acro{OSD}{ordered statistic decoding}
\end{acronym}

\begin{abstract}
Although iterative decoding of polar codes has recently made huge progress based on the idea of permuted factor graphs, it still suffers from a non-negligible performance degradation when compared to state-of-the-art \ac{CA-SCL} decoding. In this work, we show that iterative decoding of polar codes based on the \ac{BPL} algorithm can approach the error-rate performance of \ac{CA-SCL} decoding and, thus, can be efficiently used for decoding the standardized 5G polar codes. Rather than only utilizing the \ac{CRC} as a stopping condition (i.e., for error-detection), we also aim to benefit from the error-correction capabilities of the outer \ac{CRC} code. For this, we develop two distinct soft-decision \ac{CRC} decoding algorithms: a \ac{BCJR}-based approach and a \ac{SPA}-based approach. Further, an optimized selection of permuted factor graphs is analyzed and shown to reduce the decoding complexity significantly.
Finally, we benchmark the proposed \ac{CA-BPL} to state-of-the-art 5G polar codes under \ac{CA-SCL} decoding and, thereby, showcase an error-rate performance not just close to the \ac{CA-SCL} but also close to the \ac{ML} bound as estimated by \ac{OSD}.

\end{abstract}
\acresetall

\section{Introduction} \label{sec:intro}

Polar codes \cite{ArikanMain} are the first type of channel codes which are theoretically proven to achieve channel capacity. However, this is only true when the code length tends to infinity under a low complexity \ac{SC} decoder.
Due to the recent advances in polar code design and polar decoding, short length \ac{CRC}-aided polar codes are selected as the channel code for the uplink and downlink control channel of the upcoming 5G standard \cite{polar5G2018}.
The concatenation of an outer \ac{CRC} code is important to enchance the short-length error-correcting capabilities of the polar code.
Thus, efficiently decoding short \ac{CRC}-aided polar codes in terms of error-rate performance and decoding complexity/latency is an active topic of research, in particular for \ac{URLLC}.

The state-of-the-art \ac{SCL} decoder of polar codes \cite{talvardyList} benefits significantly from the possibility of a seamless integration of the \ac{CRC} into the decoding process. For this, the \ac{CRC} is used to pick the correct candidate from the list of possible codewords. However, the \ac{SCL} algorithm suffers from unfavorable decoding latency due to its sequential decoding nature and is also not well suited for iterative detection and decoding due to the hard-output nature of the decoder (and its latency).
Another polar decoding approach is to use an iterative \ac{BP} decoder \cite{ArikanBP_original} which is inherently parallel and, thus, allows high throughput implementations \cite{ErrorFloorYes_2}\cite{cammerer_HybridGPU}.
Also, it is a \ac{SISO} decoder and, thus, is very suitable for iterative detection and decoding.
However, its main drawback is a degraded error-rate performance when compared to the \ac{CA-SCL} decoder.

Several algorithms have been proposed over the past few years to enhance the error-rate performance of iterative polar decoding algorithms. Among these was an outer graph-based code (e.g., \ac{LDPC} code \cite{BP_Siegel_Concatenating} or polar code \cite{BP_felxible}) which was (partially) augmented to the inner polar code factor graph to enhance the reliability of the semi-polarized bit-channels.
In \cite{ISWCS_Error_Floor,NA_BPL}, virtual noise of low power is added to the decoder input (i.e., channel output) to avoid falling into trapping/stopping sets which also enhances the error-rate of the \ac{BP} decoder.
Furthermore, polar codes can be decoded over different permuted factor graphs which can greatly enhance the error-rate performance. This was studied for different channels (e.g., \ac{BEC} \cite{Urbanke_chCsC_BP} and \ac{AWGN} \cite{multi_trellis}) and different decoders (e.g., \ac{BP} \cite{elkelesh2018belief} and \ac{SCL} \cite{PermutedSCL}). It is worth mentioning that similar ideas were also investigated for \ac{RM} codes \cite{PermutedSCL,PermutedRM_Ivanov}.

Iterative decoding of \ac{CRC}-aided polar codes was studied in few recent works (e.g., with only \ac{CRC} detection in \cite{multi_trellis} and with the aid of the \ac{CRC} error-correcting capabilities in \cite{doan2018neural}).
However, to the best of our knowledge, there exists no iterative decoding algorithm of polar codes which achieves (or approaches) the performance of the \ac{CRC}-aided polar codes under \ac{SCL} decoding (with a sufficiently large list size).

Belief propagation list (BPL)\acused{BPL} decoding of polar codes \cite{elkelesh2018belief} is an iterative \ac{BP}-based decoding in which a list of different \ac{BP} decoders run in parallel, each with a different polar code \ac{FG}.
It was shown that it achieves the same error-rate performance as \ac{SCL} decoding of \emph{plain} polar codes (i.e., no outer \ac{CRC} code).
In this work, we focus on solving some open problems in the \ac{BPL} decoder such as: 1.) \ac{CRC} incompatibility and 2.) reducing complexity by optimizing the factor graph selection.
We combine the idea of iterative decoding of polar codes over permuted factor graphs with \ac{SISO} decoding of the \ac{CRC}, utilizing the error-correction capabilities of the \ac{CRC} (rather than only detection of errors). Our results, surprisingly, show that it is possible to approach the \ac{CRC}-aided \ac{SCL} error-rate performance with a \ac{BP}-based decoding framework. As such, the proposed algorithm suggests a competitive decoding to \ac{CA-SCL} for \ac{CRC}-aided polar codes.

\section{Polar Codes and Iterative Decoding} \label{sec:pcid}

Polar codes are based on the channel polarization concept, where $N$ synthesized channels show a polarization behavior (\emph{good} and \emph{bad} channels). Practically, polar codes can be seen as a wide range of codes, each characterized by a set of good bit-channels or its complementary set of bad channels (denoted as the information set $\mathbb{A}$ and the frozen set $\mathbb{A}_c$, respectively). 

The polarization matrix of polar codes of block size $N = 2^n$ is given by 
$\mathbf{G}_N = \mathbf{F}^{\otimes n}$, where $ \mathbf{F} = \left[ \begin{array}{ll} 1 & 0 \\ 1 & 1 \end{array}\right]$ and $\mathbf{F}^{\otimes n}$ denotes the $n$-th Kronecker power of $\mathbf{F}$. The $k\times N$ polar code generator matrix  $\mathbf{G}$ is a sub-matrix of $\mathbf{G}_N$ with row indices corresponding to $\mathbb{A}$.\footnote{Based on this insight, \ac{RM} codes can be easily related to polar codes \cite{Urbanke_chCsC_BP}.} Throughout this work, we use the notation $\mathcal{P}(N,k)$ to denote a polar code of length $N$ and code dimension $k$.

\subsection {Belief Propagation Decoding}
Unlike the \ac{SC}-based polar decoders, an iterative \ac{BP} decoding scheme \cite{ArikanBP_original} is based on the idea of message passing  over the encoding \ac{FG}, shown in Fig. \ref{fig:bp_crc} (rightmost box). Finally, a hard decision is applied on the resultant \ac{LLR} values on the \emph{left} or the \emph{right} of the \ac{FG} in order to recover the information bits (in the vector $\hat{\mathbf{u}}$) or the transmitted codeword $\hat{\mathbf{x}}$, respectively.

The conventional \ac{BP} decoder, $\text{BP}\left(N_\mathrm{it,max}\right)$,  terminates when a maximum number of iterations $N_\mathrm{it,max}$ is reached. Early stopping conditions can  speed up the decoding process \cite{earlyStop}:
\begin{itemize}
	\item $\mathbf{G}$-based: decoding terminates when $\hat{\mathbf{x}}=\hat{\mathbf{u}}\cdot \Gm_N$.
	\item \ac{CRC}-based: decoding terminates when the checks of the concatenated \ac{CRC} code are satisfied. It is important to keep in mind that in this case, the \ac{CRC} code only acts as a \emph{stopping condition} (i.e., error-detection).
\end{itemize}

\subsection{Belief Propagation List (BPL) Decoding}
\begin{figure}[t]
	\includegraphics{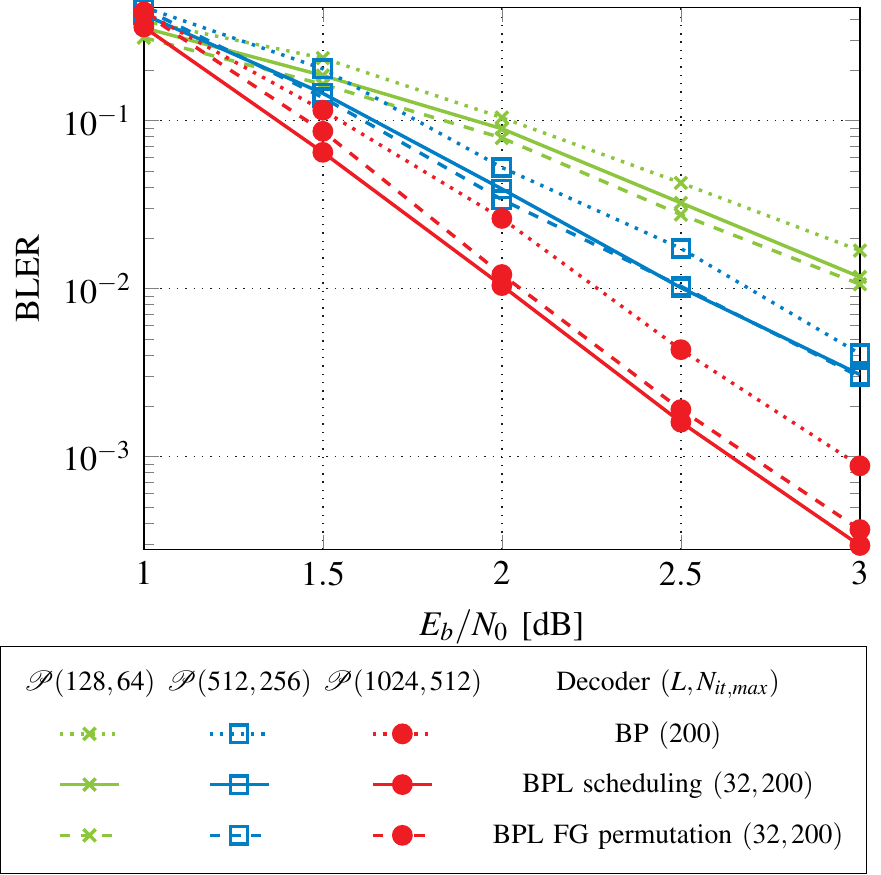}
	\vspace{-0.5cm}
	\caption{\footnotesize BLER comparison between BPL with permuted scheduling and permuted factor graphs (FGs); 5G polar codes of rate $0.5$ (i.e., $\mathcal{P}(N,N/2)$-codes); no \ac{CRC} is used.}
	\label{fig:sched_vs_perm}
	\vspace{-0.6cm}
\end{figure}

It has been observed that the stages of the encoding graph of a polar code of length $2^n$ can be permuted leading to $n!$ different graphs with the same encoding behavior \cite{Urbanke_chCsC_BP}. This enables the possibility of \ac{BP} decoding on different realizations of the \ac{FG}. Seeing that different permutations contain different loops, one \ac{FG} permutation may be better than another permutation for a specific input (i.e., transmitted codeword plus noise realization). 
\ac{BPL} exploits this idea by using a set $ \mathbb{S} $ of $L$ parallel independent \ac{BP} decoders each with a different permutation. Iterative decoding is conducted on the $L$ decoders in parallel with $N_\mathrm{it,max}$ iterations per decoder and only those codewords satisfying the $\bf{G}$-matrix-based stopping condition are declared as the set of valid polar codewords. 
Finally, that valid codeword from the $L$ parallel \ac{BP} decoders which is closest, in terms of Euclidean distance, to  the channel output $\mathbf{y}$ is picked to be the \ac{BPL} decoder output. 
Throughout this work, we use the notation BPL $\left(L,N_\mathrm{it,max}\right)$ to denote this decoder.
For more details, we refer the interested reader to \cite{elkelesh2018belief}. 

Interestingly, it was shown in \cite{Doan_2018_Permuted_BP} that a stage-wise permutation on the \ac{FG} corresponds to certain bit-wise permutations of the vectors $\mathbf{u}$ and $\mathbf{x}$. This enables permuted-decoding on the same \ac{FG} while only doing these bit-mappings. It also enabled the same permuted-decoder concept with other decoders without having to change the decoder structure itself which is highly useful for hardware implementations (see \cite{PermutedSCL}).

The selection of permutation sets $\mathbb{S}$ was originally done in a random manner \cite{multi_trellis}. Later, it was shown that a \emph{smart} permutations selection would introduce additional performance gains \cite{Doan_2018_Permuted_BP, ListFusion, PartialPermutation, BernaThesis}. However, an optimal \ac{FG} selection strategy remains an open problem.

\subsection{Factor Graph Permutations vs. Decoding Schedule Permutations}

Two implementation strategies have been found to realize the \ac{BPL} decoding of polar codes:
\begin{enumerate}
	\item Graph permutations, as originally introduced in \cite{elkelesh2018belief} where the \ac{FG} stages are permuted according to a specific permutation. Afterwards, conventional \ac{BP} decoding is conducted on the permuted factor graph, following the conventional decoding schedule (i.e., permuted stages from 1 to $n$, or $n$ to 1, are updated sequentially).
	\item Decoding schedule permutations, where conventional \ac{BP} decoding is conducted on the conventional factor graph, while permuting the decoding schedule (i.e., permuting the order in which stages are being activated/updated).
\end{enumerate}

Fig. \ref{fig:sched_vs_perm} depicts that both strategies have a similar error-rate performance, for different code lengths. We empirically observed that the factor graph permutation-based \ac{BPL} yields a better performance in the short block length regime, while its scheduling-based counterpart performs better for longer code lengths.

\section{CRC-Aided BPL Decoding} \label{sec:cabpl}
None of the previously described \ac{BPL} decoding schemes utilizes the presence of an outer \ac{CRC} code during decoding. In this paper, we propose a \ac{CA-BPL} decoding scheme that extends the standard \ac{BPL} decoding. Its underlying idea is to utilize estimated soft-information from the \ac{CRC} code to further enhance the \ac{BPL} decoder (i.e., we treat the \ac{CRC} as an outer code component).
Accordingly, we present two \ac{SISO} decoding algorithms for the \ac{CRC} code that enhance the error-correcting capability of the \ac{BPL} decoder.

\ac{CRC} codewords are defined as polynomials of degree less than $ n_{\mathrm{CRC}} $ that are divisible by the generator polynomial $ g(x) $ of degree $ r $. To check divisibility, a \ac{LFSR} of length $ r $ with input $ x_i $ and feedback according to $ g(x) $ can be used. The \ac{CRC} code can be visualized in a trellis diagram with the state $ s $ of the shift register on the vertical axis and the discrete time index $ i $ on the horizontal axis. 
Fig. \ref{fig:crc_lfsr_trellis} shows the \ac{LFSR} and the trellis of a CRC-2 code with polynomial $g(x)~=~x^2~+~x~+~1$ as an example. The register is initialized to the all-zero state. Every time index, a state transition $ (s',s)$ from state $ s' $ to $ s $ occurs, depending on the input bit value $ x_i $. A bit sequence fulfills the \ac{CRC}, i.e., it is a codeword, if and only if the shift register returns to the all-zero state when the last bit enters the circuit. For all transitions, $ \lambda_{\mathrm{trans}}(s',s) \in \{0,1\} $ denotes the bit value of the transition. Likewise, $ U_0 $ and $ U_1 $, depicted as solid black and dashed red lines in the trellis diagram,  denote the set of edges $ (s',s) $ corresponding to the 0 and 1 transitions, respectively.  

\begin{figure}[t]
	\centering
	\begin{subfigure}{.4\columnwidth}
		\centering
		\resizebox{\columnwidth}{!}{
			\includegraphics{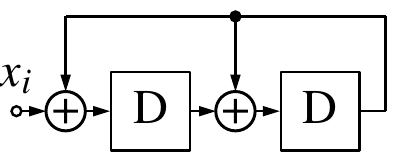}
		}
	\end{subfigure}%
	\begin{subfigure}{.6\columnwidth}
		\centering
		\resizebox{\columnwidth}{!}{
			\includegraphics{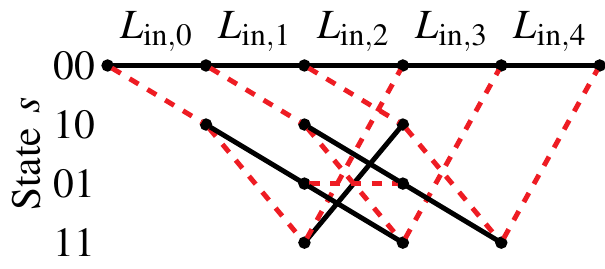}
		}
	\end{subfigure}
	\caption{\footnotesize Linear feedback shift register (LFSR) and length $ n_{\mathrm{CRC}}=5 $ trellis of the \ac{CRC} code with $g(x)~=~x^2~+~x~+~1$.}
	\label{fig:crc_lfsr_trellis}
\end{figure}

We use the \ac{BCJR} algorithm \cite{BCJR} to enable \ac{MAP} decoding of the \ac{CRC} code. Thus, we  briefly present the log-domain \ac{BCJR} algorithm \cite{shuLin}. 

\begin{align}\label{eq:bcjr_fp}
\tilde{\alpha}_{i+1}(s) = \maxstar_{s'}\left( \tilde{\alpha}_{i}(s')   - L_{\mathrm{in},i}\cdot \lambda_{\mathrm{trans}}(s',s) \right)
\end{align}
$ \forall s $ and $ i = 0, ..., n_{\mathrm{CRC}}-1 $ starting with $ \tilde{\alpha}_{0}(0)~=~0 $.

\begin{align}\label{eq:bcjr_bp}
\tilde{\beta}_{i}(s') = \maxstar_{s}\left( \tilde{\beta}_{i+1}(s)   - L_{\mathrm{in},i}\cdot \lambda_{\mathrm{trans}}(s',s) \right)
\end{align}
$ \forall s' $ and $ i = n_{\mathrm{CRC}}-1, ..., 0 $ starting with $ \tilde{\beta}_{n_\mathrm{CRC}}(0)~=~0 $.

\begin{align}\label{eq:bcjr_ext}
L_{\mathrm{out},i} = & \maxstar_{U_0} \left(  \tilde{\alpha}_{i}(s') + \tilde{\beta}_{i+1}(s) \right) 
- \maxstar_{U_1} \left(  \tilde{\alpha}_{i}(s') + \tilde{\beta}_{i+1}(s) \right).
\end{align}

\begin{figure}[t]
	\centering
	\begin{subfigure}{.5\columnwidth}
		\centering
		\resizebox{\columnwidth}{!}{
			\includegraphics{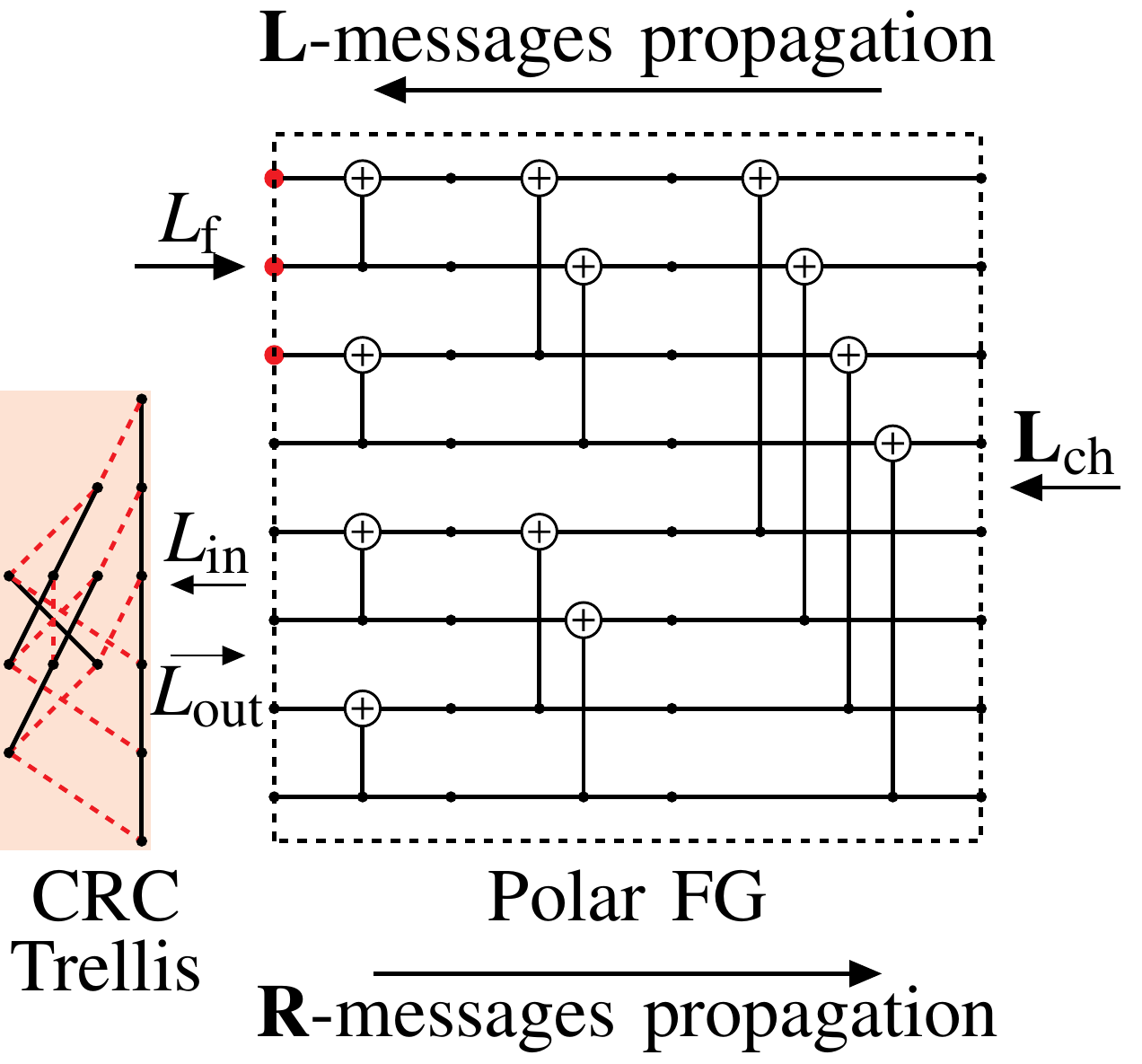}
		}
	\end{subfigure}%
	\begin{subfigure}{.5\columnwidth}
		\centering
		\resizebox{\columnwidth}{!}{
			\includegraphics{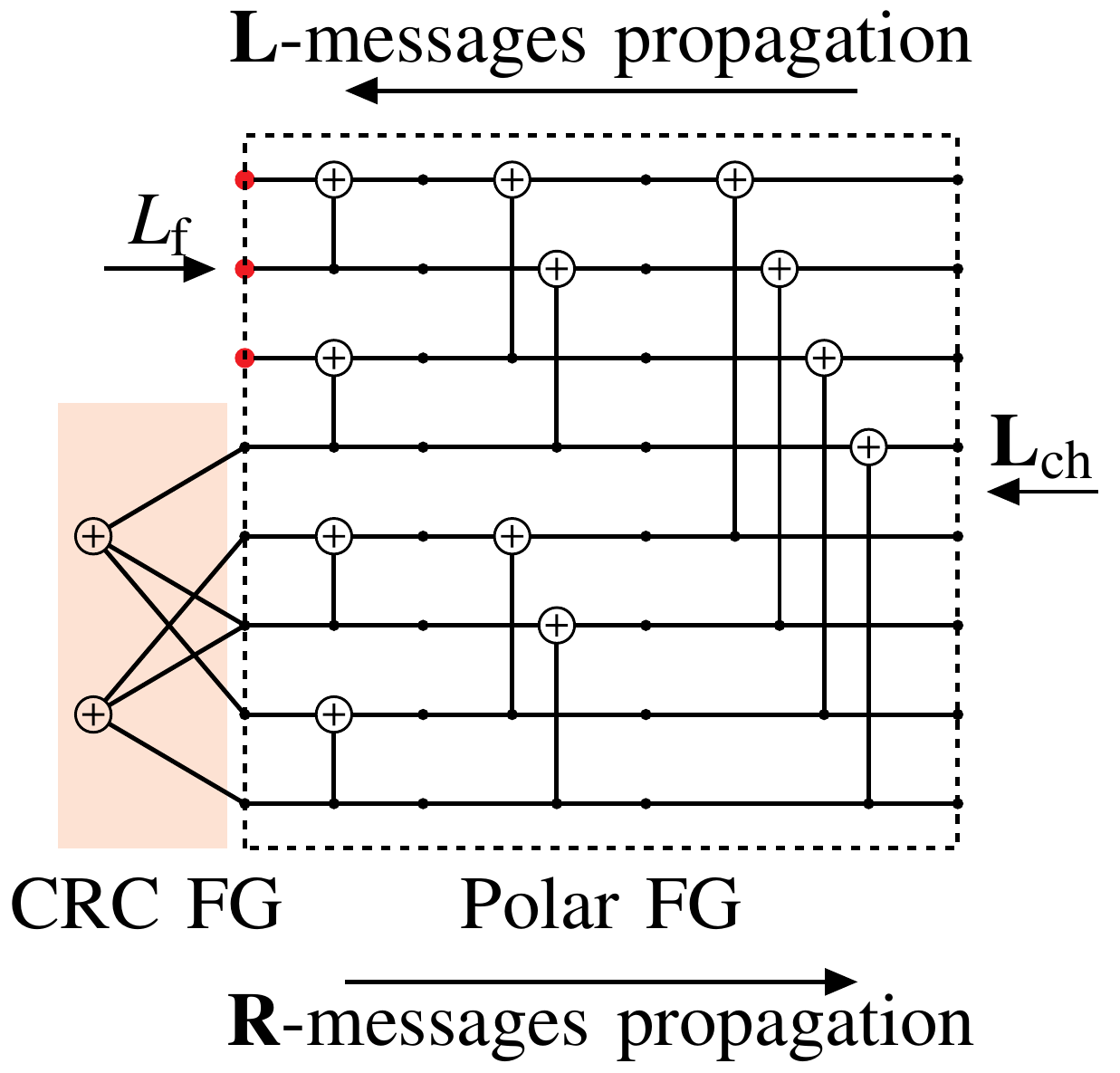}
		}
	\end{subfigure}
	\caption{\footnotesize Polar + \ac{CRC} decoders based on the polar factor graph (FG) augmented with BCJR trellis (left) and SPA factor graph (right) for the CRC. The code used is a $\mathcal{P}(8,5)$ polar code concatenated with a CRC code with polynomial $g(x)~=~x^2~+~x~+~1$.
	}
	\label{fig:bp_crc}
	\vspace{-0.6cm}
\end{figure}

The inputs to the \ac{BCJR} algorithm are the \acp{LLR} $ L_{\mathrm{in},i} $ of each bit of the noisy \ac{CRC} codeword. The \ac{BCJR} algorithm estimates \aposteriori log-probabilities $ \tilde{\alpha} $ and $ \tilde{\beta} $ recursively, based on the current state along with state transition probabilities in a forward pass (\ref{eq:bcjr_fp}) and a backward pass (\ref{eq:bcjr_bp}) on the trellis. The estimates are combined in a final step (\ref{eq:bcjr_ext}) to obtain the extrinsic L-value $ L_{\mathrm{out},i} $ for each bit. To compute the sum of probabilities in the log-domain, we use the $ \maxstar $ operator, which can be expressed as $ \maxstar\limits_{j}(x_j)~=~\log\left( \sum\limits_{j}{ e^{x_j}}\right) $.

From the equations it is obvious that the complexity of the \ac{BCJR} algorithm scales proportional to the number of states in the underlying state machine. Therefore, the overall complexity is $ \mathcal{O}(n_{\mathrm{CRC}} \cdot 2^{r}) $, making this algorithm infeasible for long \ac{CRC} lengths $ r $.

To counteract complexity, we propose a second \ac{SISO} decoder. It is based on the \ac{SPA} used to decode \ac{LDPC} codes. \ac{SPA} operates on the parity-check matrix $ \Hm_{\mathrm{CRC}} $ of the \ac{CRC} code. For details on the \ac{SPA}, we refer the interested reader to \cite{shuLin}.
$ \Hm_{\mathrm{CRC}} $ is directly obtained from the systematic generator matrix, which is generated by the \ac{CRC} codewords corresponding to the unit vectors. As the \ac{SPA} is known to have poor performance on high density parity-check matrices, the density of $ \Hm_{\mathrm{CRC}} $ is reduced using row operations. As a heuristic, a greedy algorithm that iteratively combines two rows and replaces the one with higher Hamming weight, proved to be sufficient for this task.\footnote{We empirically observed that the exact form of $ \Hm_{\mathrm{CRC}} $ has only minor impact on the error-rate performance.}

For the example \ac{CRC} code with polynomial $g(x)~=~x^2~+~x~+~1$, the resulting parity-check matrix is

\begin{align}
\Hm_{\mathrm{CRC}}  = \left[ \begin{array}{rrrrr}
1 & 0 & 1 & 1 & 0 \\
0 & 1 & 1 & 0 & 1 \\
\end{array}\right]. \nonumber
\end{align}

Fig. \ref{fig:bp_crc} shows how both the BCJR and SPA decoders of the \ac{CRC} code are augmented to the left side of the polar decoder. The \ac{CRC} decoder only interacts with the non-frozen bit-positions $ \mathbb{A} $ of the polar code. After each right-to-left message propagation on the polar \ac{FG}, the \ac{CRC} decoder is fed with the left-most L-messages of the information bit channels $ \Lm_{\mathrm{in}} = \Lm_{0,\mathbb{A}}$. Then, the \ac{CRC} decoder is updated and its output is fed back into the left-most R-message $ \Rm_{0,\mathbb{A}} = \Lm_{\mathrm{out}}$, followed by a left-to-right message update on the polar \ac{FG}. 

In \ac{CA-BPL}, a total of $ L $ permuted factor graphs is used, each with its own instance of the \ac{CRC} decoder. Throughout this work, we only consider factor graph permutation-based \ac{BPL} rather than scheduling-based \ac{BPL} (cf. Fig.~\ref{fig:sched_vs_perm}).

\section{Results} \label{sec:results}
We evaluate the performance of the \ac{CA-BPL} algorithm on a concatenated polar-\ac{CRC} code with codelength $ N=128 $ and code dimension $ k=64 $. As CRC polynomial, CRC-6 from the 5G standard \cite{polar5G2018} $ g(x)~=~x^6~+~x^5~+~1 $ is used. Consequently, the polar code has 70 non-frozen bits, to account for the rate loss of the outer CRC code. For the sake of reproducibility, we use the bit-reliability order specified in the 5G standard \cite{polar5G2018} as the design criterion of the inner polar code (i.e., $\mathbb{A}$).

\subsection{Permutation Selection} \label{sub:perm}

\begin{figure}[t]
	\includegraphics{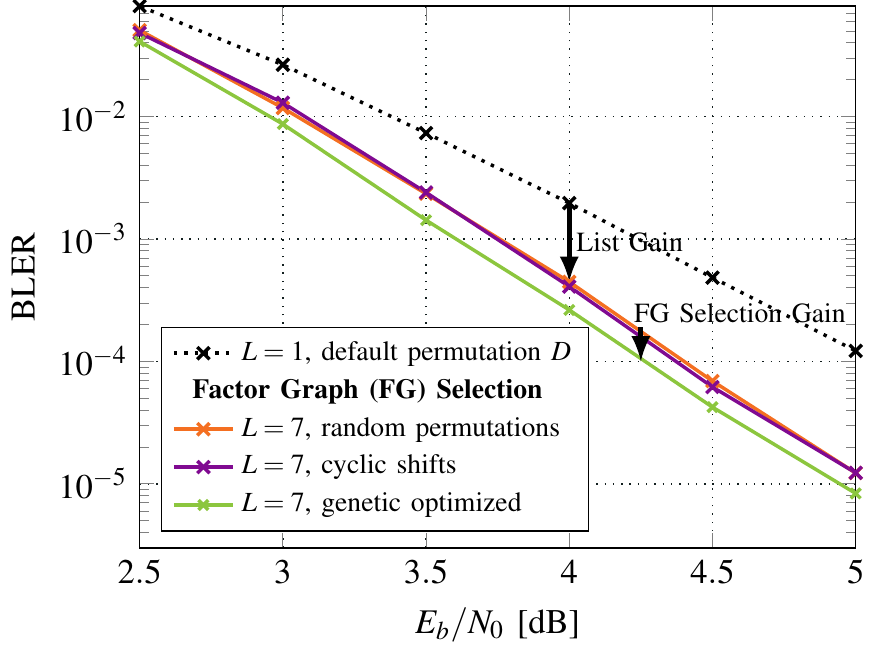}
	\vspace{-0.7cm}
	\caption{\footnotesize BLER comparison between CRC-aided BPL with different permutation selection methods; BCJR-based CRC decoders were used and $N_{\mathrm{it,max}} =200$. The considered CRC-aided polar code has a code length $N=128$, code dimension $k=64$ and CRC-6 is used, thus, the overall code rate is $0.5$.}
	\label{fig:list_gain}
	\vspace{-0.6cm}	
\end{figure}
One possible approach to reduce the complexity of the proposed decoding scheme would be to \emph{smartly} select the \ac{FG} permutations reducing the number of required \ac{FG}s.

Remember that, in \ac{BPL}, as long as the received vector is \ac{ML}-decodable, only one of the $ L $ permuted \ac{BP} decoders has to converge to the correct codeword to successfully decode. In other words, the block error probability of the \ac{ML}-decodable received sequences $ P\left(E_{\mathrm{BPL}} \middle| \overline{E_\mathrm{ML}} \right) $ of the \ac{BPL} decoder is the probability of all the constituent \ac{BP} decoders being in error. This is expressed as

\begin{align}\label{eq:bpl_error}
P\left(E_{\mathrm{BPL}(\mathbb{S})} \middle| \overline{E_\mathrm{ML}} \right) =  P\left(\bigcap_{i \in \mathbb{S}}{E_{i}} \middle| \overline{E_\mathrm{ML}} \right)
\end{align}
where $ \overline{E_\mathrm{ML}} $ denotes an \ac{ML} decoder success event, $ E_{\mathrm{BPL}} $ denotes the event of the \ac{BPL} decoder block error, and $ E_i $ denotes the event of the \ac{BP} decoder using \ac{FG} permutation $ i $ not converging to the correct codeword within $N_{\mathrm{it,max}} $ iterations.

Therefore, to use the full potential of \ac{BPL}, it is necessary to find a set $ \mathbb{S} $ of $ L $ factor graph permutations that yields the best joint \ac{BLER} performance. The intuition behind this idea is to have more \emph{diverse} permutations that lead to more differentiated results (i.e., diversity in the final list of candidates) than multiple good permutations that lead to the same results (i.e., having the same  candidate multiple times in the list). 
Thus, we would like to minimize $ P\left(E_{\mathrm{BPL}(\mathbb{S})} \middle| \overline{E_\mathrm{ML}} \right) $ over all possible sets $ \mathbb{S} $ with the constraint $ |\mathbb{S}|=L $

\begin{subequations}
\begin{align}
&\underset{\mathbb{S}}{\text{minimize}}
&& P\left(\bigcap_{i \in \mathbb{S}}{E_{i}}\middle| \overline{E_\mathrm{ML}} \right) \nonumber \\
&\text{subject to}
&& |\mathbb{S}|=L. \nonumber
\end{align}
\end{subequations}

For a length $ N=128 $ code, there exist $ 7!=5040 $ different permutations of the \ac{FG}. Thus, the search space of the optimization problem contains $ {5040 \choose L} $ possible permutation sets. As we can only estimate the probabilities using Monte Carlo simulation and error events are comparably rare, naive optimization is computationally infeasible. For the 5G construction of the polar code, however, we empirically observe that, regarding standalone \ac{BP} decoders, the default permutation $ D $ achieves the lowest \ac{BLER}. By assuming $ D $ is in the optimum set $ \mathbb{S} $ we can factor equation (\ref{eq:bpl_error}) into a conditional \ac{BP} \ac{BLER} and a term corresponding to the \emph{list gain}:

\begin{align}
P\left(\bigcap_{i \in \mathbb{S}}{E_{i} }  \middle| \overline{E_\mathrm{ML}} \right) \stackrel{D \in \mathbb{S}}{=}  P\left(E_D \middle| \overline{E_\mathrm{ML}}\right) \cdot \underbrace{P\left(\bigcap_{i \in \mathbb{S} \setminus \{D\}}{E_{i}} \; \middle| \; E_{D}, \overline{E_\mathrm{ML}}  \right)}_{\text{List Gain}}. \nonumber
\end{align}

This factorization simplifies the optimization greatly, as $ P\left(E_{i} \; \middle| \; E_{D}, \overline{E_\mathrm{ML}} \right) \gg P\left(E_{i}  \middle| \overline{E_\mathrm{ML}} \right) $, due to the correlation of the permuted \ac{FG} decoders. Consequently, a comparably small dataset of $2\cdot10^4$ samples could be used to find the remaining optimal $ L-1 $ other permutations. The collected dataset contains $ \yv $ vectors at $ E_{\mathrm{b}}/N_0 = 4 \text{ dB} $ for which the \ac{BP} decoder, using the default permutation $D$, fails to decode. We then evaluate the convergence of all other 5039 permutations. Finally, a genetic algorithm is applied on 15000 samples of the dataset to find the optimal $ \mathbb{S} $ by minimizing the estimate of $ P\left(\bigcap_{i \in \mathbb{S} \setminus \{D\}}{E_{i}} \;  \middle| \;  E_{D} \right) $, i.e., maximizing the list gain. The last 5000 samples serve as a validation set to test whether the solution generalizes well. In Fig. \ref{fig:list_gain} we compare different permutation selection methods for \ac{CA-BPL} with list size $ L=7 $ based on \ac{BLER} performance. The genetic optimized list outperforms both random permutations and the 7 cyclic shifts of the default permutation.

\subsection{Error-Rate Performance} \label{sub:sim}
\begin{figure}[t]
	\includegraphics{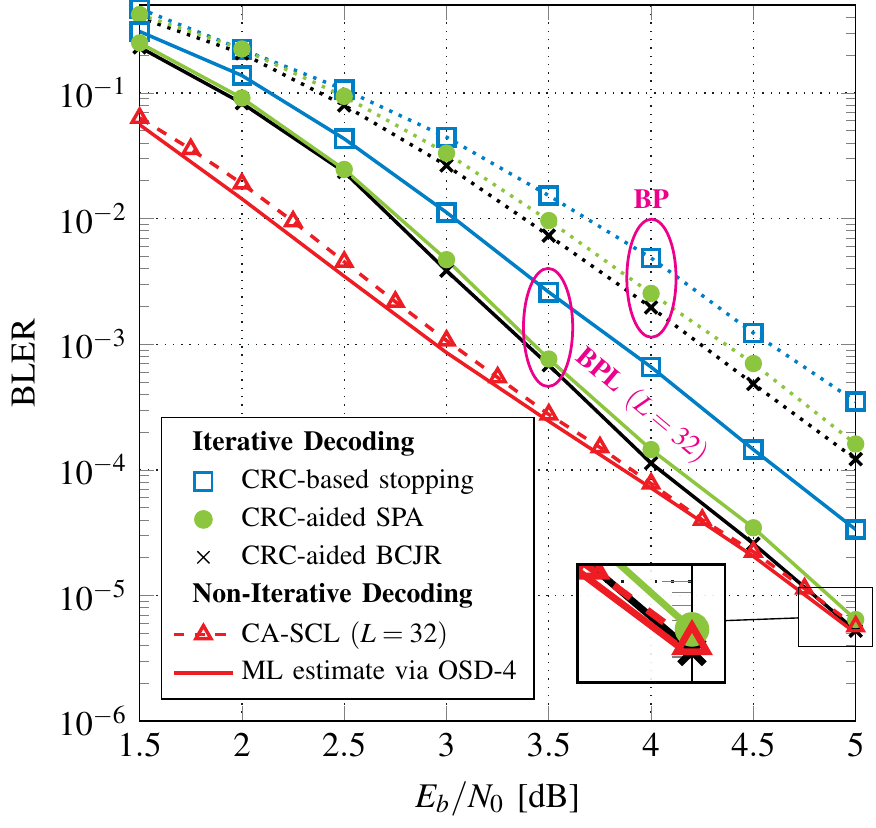}
	\vspace{-0.7cm}
	\caption{\footnotesize BLER comparison between different CRC-aided BPL decoders and CRC-aided SCL decoder; the considered CRC-aided polar code has a code length $N=128$, code dimension $k=64$ and CRC-6 is used, thus, the overall code rate is $0.5$. The iterative decoders use a maximum number of iterations of $ N_{\mathrm{it,max}} = 200$.
	}
	\label{fig:CA-BPL}
	\vspace{-0.6cm}
\end{figure}

We compare a classical \ac{BPL} decoder that uses the \ac{CRC} solely as a stopping condition to our proposed \ac{CA-BPL} decoding scheme, with both \ac{BCJR} and \ac{SPA} \ac{CRC} decoders. Additionally, we use \ac{CA-SCL} decoding with list size $ L=32 $ and order-4 \ac{OSD} \cite{OSD} as a lower bound on the \ac{BLER} performance. The \ac{CRC} parity-check ensemble used has (edge-perspective) variable node and check node degree distribution polynomials $  \lambda(Z)~=~0.06~+~0.16Z^1~+~0.3Z^2~+~0.3Z^3~+~0.15Z^4~+~0.03Z^5 $  and $ \rho(Z)~=~0.66Z^{32}~+~0.34Z^{33} $, respectively.

The \ac{BPL} and \ac{CA-BPL} decoders use $ N_{\mathrm{it,max}} = 200$ iterations and $ L=32 $ parallel decoders. The \ac{FG} permutations were selected using the optimization technique described above. Fig.~\ref{fig:CA-BPL} shows the \ac{BLER} performance of the decoders over the \ac{AWGN} channel using \ac{BPSK} modulation. The proposed \ac{CA-BPL} scheme outperforms a classical \ac{BPL} (with CRC used as a stopping criterion) by 0.5~dB at a \ac{BLER} of $10^{-3}$. Note that, this gain is solely due to benefiting from the CRC error-correction capabilities rather than only detecting errors. It is worth mentioning that \ac{CA-BPL} approaches the performance of \ac{CA-SCL} closer than 0.15 dB at a \ac{BLER} of $10^{-4}$. The same \ac{BLER} performance as \ac{CA-SCL} is reached at $ E_{\mathrm{b}}/N_0 = 5 \text{ dB} $.  It is quite remarkable that the much simpler \ac{SPA}-aided \ac{BPL} decoder is only marginally worse than the (optimal) \ac{BCJR}-aided \ac{BPL}.

\section{Conclusion} \label{sec:conc}
We have introduced a \ac{CRC}-aided extension of the iterative \ac{BPL} decoding algorithm, where the \ac{CRC} is used for error-correction rather than the plain error-detection. For this, we have developed two soft-in/soft-out decoding techniques for \ac{CRC} codes based on the trellis-based \ac{BCJR} algorithm and the iterative \ac{SPA} decoder.
To account for the additional decoding overhead, we have further optimized the selection of factor graph permutations and have shown that the required list size can be reduced.
As a result, the proposed \ac{CA-BPL} algorithm can compete with \ac{CA-SCL} decoding in the typical SNR region of interest while promising lower decoding latency and offering the potential for parallel algorithm implementations. For a large enough list size, the proposed algorithm even approaches the estimated \ac{ML} performance of the concatenated polar+\ac{CRC} coding scheme. To the best of our knowledge, these are the best iterative polar+CRC decoding results that have been reported in literature.
Although we have given some intuition why the selection of factor graph permutation works, we leave it open for future work to derive a theoretical analysis of \emph{why} a certain set of permutations yields a better overall performance than other selections.

\bibliographystyle{IEEEtran}
\bibliography{references}
\end{NoHyper}
\end{document}